\documentclass[iop, twocolumn, tighten]{emulateapj}
\newcommand{\NAME}{Suzaku J1552$+$2739\,\,}
\newcommand{\PROGNAME}{DisPerSE\,\,}
\newcommand{\BCG}{2MASX J15525730$+$2739474\,\,}

\slugcomment{The Astrophysical Journal Letter, accepted}
\shortauthors{Kawahara et al.}
\shorttitle{A New Merging Group at a Filamentary Junction}

\begin{document}

\title{{\it SUZAKU} Observation of a New Merging Group of Galaxies at a Filamentary Junction}

\author{Hajime Kawahara\altaffilmark{1}, Hiroshi Yoshitake\altaffilmark{2,3},  Takahiro Nishimichi\altaffilmark{3} and Thierry Sousbie\altaffilmark{3}} 
\altaffiltext{1}{Department of Physics, Tokyo Metropolitan University,
  Hachioji, Tokyo 192-0397, Japan}
\altaffiltext{2}{Institute of Space and Astronautical Science, Japan Aerospace Exploration Agency (ISAS/JAXA), Kanagawa 229-8510, Japan}
\altaffiltext{3}{Department of Physics, The University of Tokyo, 
Tokyo 113-0033, Japan}
\email{kawa\_h@tmu.ac.jp}

\begin{abstract}
We report on a new merging group of galaxies, \NAME at $z \sim 0.08$ revealed by a {\it SUZAKU} observation. The group was found by observing a junction of galaxy filaments optically identified in the Sloan Digital Sky Survey spectroscopic data. \NAME exhibits an irregular morphology and presents several peaks in its X-ray image. A bright elliptical galaxy, observable in the central peak, allows the localization of the group at $z=0.083$. We found a significant hot spot visible in the X-ray hardness map, close to the second peak. The spectroscopic temperature is $T=1.6^{+0.4}_{-0.1}$ keV within $ R_{500} = 0.6 \, \mathrm{Mpc}$ and  $T = 3 - 5 $ keV in the hot spot. We interpret those results as \NAME being located in the center of a major merging process. The observation of a galaxy group showing multiple X-ray peaks and a hot spot at the same time is rare and we believe in particular that the study of \NAME potentially presents a significant interest to better understand the dynamical and thermal evolution of the intragroup and intracluster medium, as well as its relation with surrounding environment. 
\end{abstract}

\section{Introduction}
The filamentary structures revealed by large scale galaxy surveys form the so-called cosmic web \citep{1996Natur.380..603B}, that traces the matter and gas distributions in the Universe. The theory behind the formation of the cosmic web is now well established \citep[see e.g.][]{1996Natur.380..603B,1996AAS...189.1303P}: the rare high density peaks that exist within the primordial Gaussian random distribution of matter latter collapse into clusters, while the primordial tidal fields they create drive the evolution of the filaments, already imprinted within the topology of the initial distribution. In a sense, filaments form the highways along which matter is transported from the lower density regions that will latter form the voids to those high density peaks, located at their junctions.

Because of the strong gravitational potential created by the dark matter filaments, the galaxy filaments are very similar to them. Hence, if one is able to identify them within the galaxy distribution, they can be used to search for undetected gas components. For instance,  X-ray absorption lines across large filaments were interpreted as the signal from the missing baryons \citep{2009ApJ...695.1351B,2010arXiv1008.5148W}. Massive X-ray clusters are also often discovered within large galaxy filaments or at the intersection of two or more filaments \citep[e.g.][]{2000A&A...355..461A,2004A&A...416..839B,2004A&A...425..429C,2007A&A...470..425B,2008A&A...491..379G}.

As mentioned previously, clusters form at the junction of filaments, and it therefore also seems reasonable to search for new X-ray haloes precisely at those filamentary junctions of galaxies. In this paper, we show the results of an X-ray observation with {\it SUZAKU} \citep{2007PASJ...59S...1M} at a filamentary junction, identified within the Sloan Digital Sky Survey (SDSS), and where no X-ray detection was previously achieved. We report the detection of a new galaxy group with significant merging features. Throughout the paper, we assume a $\Lambda$ CDM universe with $\Omega_0=0.27$, $\Omega_\Lambda=0.73$, and $h=0.71$.

\section{Target Selection}

\begin{figure*}[!tbh]
  \begin{center}
\includegraphics[height=80mm]{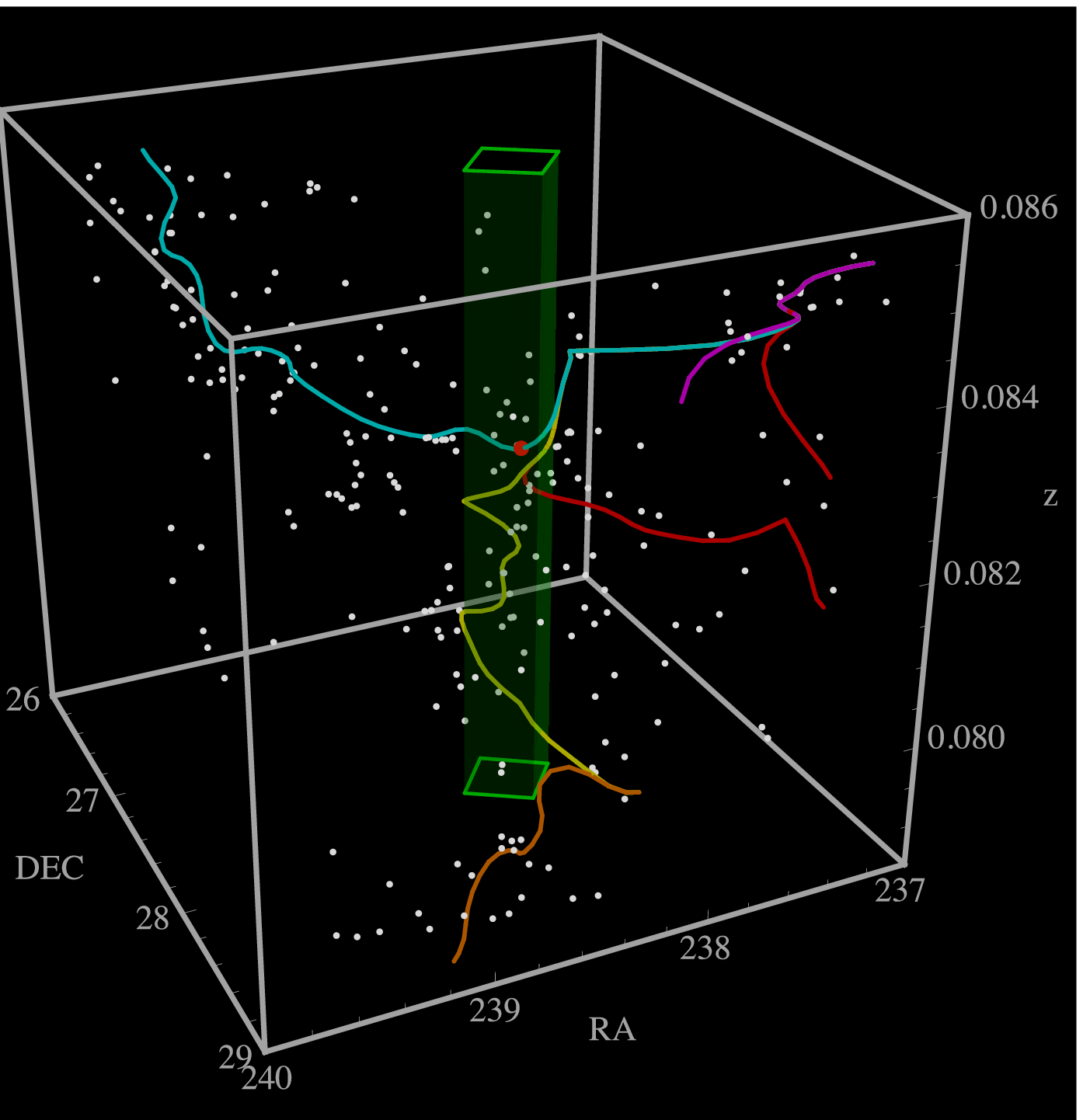}
\includegraphics[height=80mm]{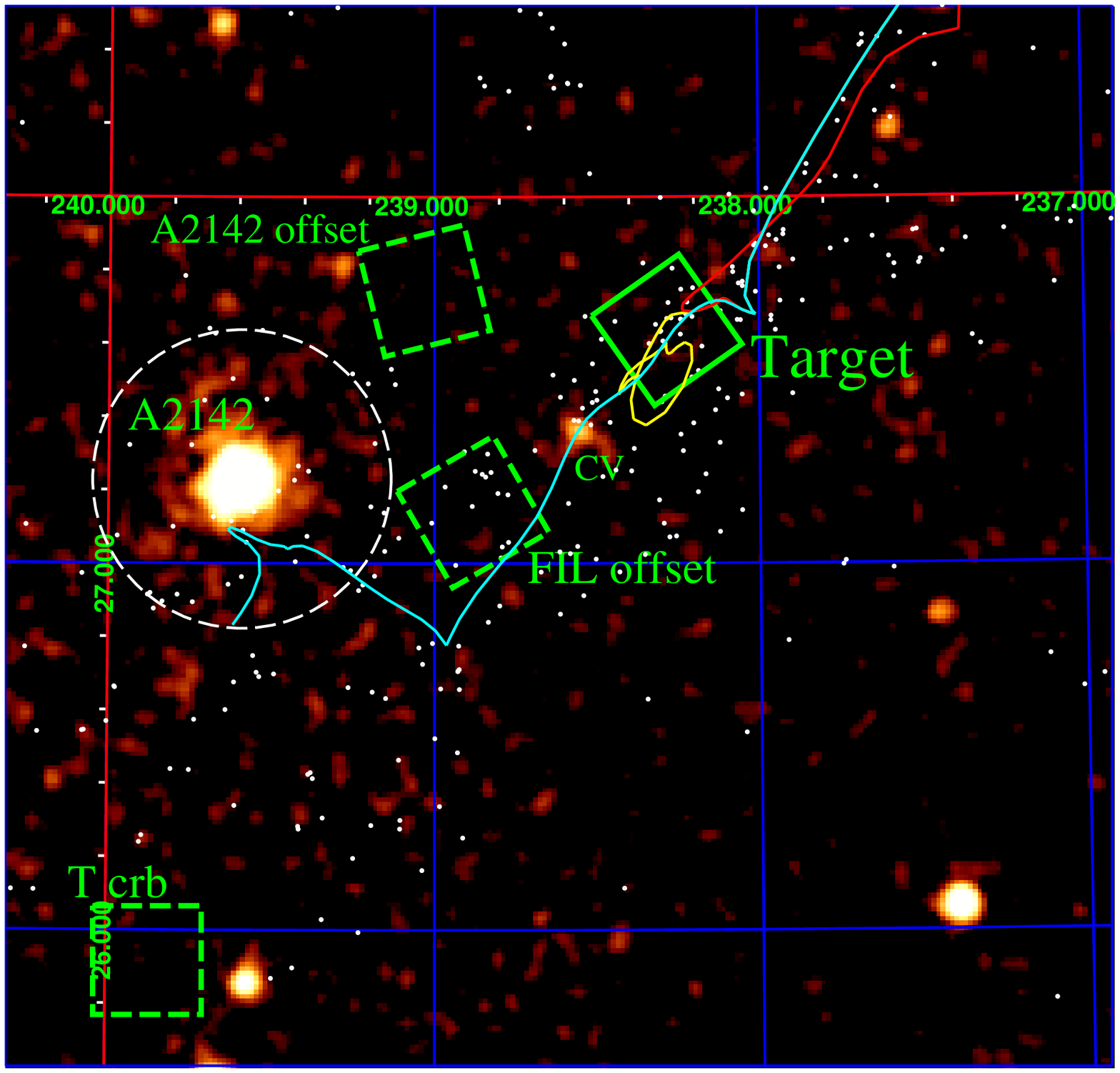}
  \end{center}
  \caption{Configuration of the target and the galaxy distribution.  Left: The SDSS spectroscopic galaxies (dots) around the observed region painted by green. Solid curves display filamentary structure (skeleton) computed by \PROGNAME.  A red point indicates the BCG, \BCG . Right: The target overlaid on the RASS 0.1-2.4 keV map (color) and the SDSS spectroscopic selected galaxies at $z=0.080-0.086$ (dots). The main target of the paper and three offset regions are shown by solid and dashed squares. The dashed circle around A2142 provides $R_{200}$ with $T=8.1$ keV \citep{2000ApJ...541..542M}.   \label{fig:pos}}
\end{figure*}

The observational target was selected to aim at the diffuse intergalactic gas within galaxy filaments and was observed by SUZAKU for 80 ksec from July 31 in 2010. The identification of the correct environment was achieved using the New York University Value-Added Galaxy Catalogue \citep{2005AJ....129.2562B,2008ApJ...674.1217P,2008ApJS..175..297A}, extracted from the spectroscopic data of the SDSS Data Release 7 (Fig. [\ref{fig:pos}]) by a visual inspection at first and was confirmed by \PROGNAME \citep{2010arXiv1009.4015S,2010arXiv1009.4014S} after the submission of the proposal. The actual 3D filamentary structure in the vicinity of our target is shown in Figure \ref{fig:pos}:  several filaments connect around redshift $z \sim 0.083$, where a bright elliptical galaxy, \BCG \citep[hereafter BCG; the brightest cluster galaxy]{2002ApJS..143....1M} is located.

As shown in the X-ray image by ROSAT All Sky Survey (RASS) displayed in Figure  \ref{fig:pos} (right), the massive cluster Abell 2142 ($z=0.09$) is situated near the target. More importantly, no X-ray source within the field of view is found in the ROSAT faint source catalog in 0.1 - 2.4 keV map \citep{2000yCat.9029....0V} \footnote[1]{http://heasarc.gsfc.nasa.gov/W3Browse/rosat/rassfsc.html} . We note that there is no corresponding object in the SDSS group catalogue with the {\it spectroscopic } dataset \citep{2010A&A...514A.102T}, while this region have been identified as a galaxy cluster candidate around $z \sim 0.1$ by previous {\it photometric} surveys: an optically selected galaxy cluster catalogue of the digitized Second Palomar Observatory Sky Survey \citep{2004AJ....128.1017L} and the maxBCG technique using the SDSS photometric data to select candidates for gravitational arcs search \citep{2007ApJ...660.1176E}. 

\section{An X-ray image and a Hardness Ratio Map}

\begin{figure*}[!tbh]
\includegraphics[width=\linewidth]{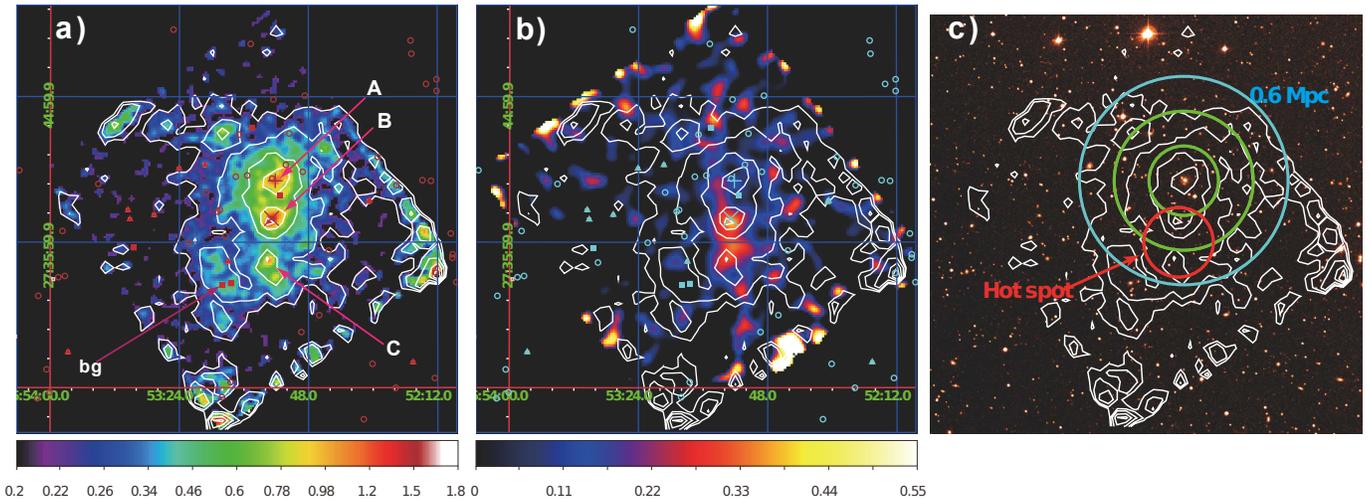}
  \caption{Left: The combined image of XIS0, 1, and 3 (0.5-5.0 keV). The vignetting is corrected for by a mono-energetic exposure map extracted at a photon energy of 1 keV. The unit is $10^{4} \, \mathrm{counts \, s^{-1} \, cm^{-2} \, sr^{-1}}$. The SDSS spectroscopic identified galaxies between $z=0.07-0.09$, $z=0.09-0.1$ and $z=0.1-0.3$ (background) are shown by circles, triangle and boxes, the central BCG at z=0.083 by a cross, and a galaxy at z=0.087 on the peak B by $\times$. Middle: the hardness ratio map of the combined image. Contour lines indicate the X-ray image shown in panel a.  Right : Spectroscopic analysis regions. Contour lines indicate the X-ray image overlaid on the photographic image provided by the DSS. \label{fig:HR}}
\end{figure*}

We analyze cleaned event files version 2.5 of the three CCD chips (XIS 0, 1, and 3) with both 5 $\times$ 5 and 3 $\times$ 3 editing and normal clocking modes after reprocessing with the standard selection criteria: the energy correction by {\it xispi}, the removal of hot/flickering pixels by {\it cleansis}. The exposure corrected image shown in Figure \ref{fig:HR}a exhibits the presence of a diffuse X-ray halo with an irregular morphology, namely, a new object \NAME. We denote three peaks as peak A, B and C as shown in Figure \ref{fig:HR}a.  The gravitational center of \NAME is likely located near the peak A because of the presence of the BCG (see also the optical image by Digitized Sky Survey (DSS) in  Fig. [\ref{fig:HR}] c  ). 

Computing the hardness ratio map shown in Figure \ref{fig:HR}b, we found ``the hot spot'' from the peaks B toward C. The definition of the hardness ratio is $\mathrm{HR \equiv (H-S)/(H+S)}$ with the hard (H) and soft (S) band images of the combined front-illuminated (FI) and back-illuminated (BI) images, where the energy range of H and S are 1.1-2.0 keV and 0.4-1.1 keV. The combination of the irregular morphology and the presence of the hot spot is a common feature in the merging cluster \citep[e.g.][]{2005A&A...432..809D,2006A&A...446..417F}. 

\section{Spectroscopic Analysis}

We now measure the spectroscopic temperature. All spectral fittings were performed with XSPEC 12.6.0 \citep{1996ASPC..101...17A} with Suzaku Calibration Database. We exclude data having a revised cutoff rigidity (COR2) less than 6 GV. Using {\it xissimarfgen}, we created two different the Ancillary Response Files (ARFs), $\mathrm{A^B}$ and  $\mathrm{A^U}$ assuming the observed XIS image in the energy range of 1-4 keV and uniform sky emission, which are applied to the group component and the cosmic X-ray background (CXB) plus other diffuse galactic components, respectively \citep{2007PASJ...59S.113I}. 

To determine the foreground and background model, we choose TCRB as a fiducial offset (Fig. [\ref{fig:pos}] (right). We also analyzed A2142 offset and FIL offset\footnote{FIL offset was obtained as part of the same observation.} as references although these two regions are close to A2142 and possibly include leakage. After excluding a central target of TCRB and several point-like sources, we simultaneously fit the spectra of the FI and BI chips by the two-temperature model: {\it phabs $\ast$ }({\it pow$+apec_1$} ){\it $+apec_2$}, where {\it pow} indicates the power law of the CXB with the photon index $\Gamma=1.41$ \citep[fix; ][]{2002PASJ...54..327K}, and the CXB intensity $S_\mathrm{CXB}$, {\it $apec_1$} is the thermal emission model with $T=T_1$ \citep[APEC;][]{2001ApJ...556L..91S} interpreted as the transabsorption emission \citep[TAE;][]{2000ApJ...543..195K}  from outside of the Galaxy , such as the Galactic Halo emission , {\it  $apec_2$ } assumes the thermal emission with $T=T_2$ from inside of the Galaxy, such as the local hot bubble and we fix the Galactic absorption ({\it phabs}) using the neutral hydrogen column density $N_{\mathrm{H}}$ provided by \citet{2005A&A...440..775K}. We use the data in the energy range of 0.4 - 5.0 keV and 0.5 - 5.0 keV for the BI and FI chips. The best fit values are given in Table \ref{tab:spec}. The CXB intensity of TCRB is consistent with the results by ASCA, $S_\mathrm{CXB,asca}=(6.38 \pm 0.07 \pm 1.05) \times 10^{-8} \, \mathrm{erg \, cm^{-2} s^{-1} sr^{-1} keV^{-1}}$ \citep[90 \% statistical and systematic error; ][]{2002PASJ...54..327K}. 

\begin{table*}[htb]
  \caption{\label{tab:spec} The spectral fitting.}
    \begin{center}
      \scalebox{1.0}[1.0]{
    \begin{tabular}{lcccccccccccc}
      \hline\hline
   \multicolumn{12}{c}{ Offset (Two temperature model$^\dagger$) } \\
   \hline
     name & OBSID & $N_\mathrm{H}$ $\,^\mathrm{a}$  & $\Gamma$ & $S_{\mathrm{CXB}} \,^\mathrm{b} $ & $kT_1$ (keV) & norm$_1$ $\,^\mathrm{c}$ & $kT_2$ (keV) & norm$_2$ $\,^\mathrm{c}$ & $\chi^2$/dof & & \\
      \hline
TCRB & 401043010 & 4.73$^\ast$ &   1.41$^\ast$  &  $5.21_{-0.19}^{+0.2}$ &  $0.286_{-0.019}^{+0.023}$ &  $0.56_{-0.08}^{+0.08}$ &  $0.087_{-0.016}^{+0.01}$ &  $4.52_{-2.0}^{+8.8}$ & 1.30 \\
FIL offset & 805029010 & 3.47$^\ast$ & 1.41$^\ast$ & $5.83_{-0.18}^{+0.17}$ &  $0.284_{-0.018}^{+0.02}$ &  $0.69_{-0.09}^{+0.10}$ &  $0.087_{-0.015}^{+0.008}$ &  $9.0_{-3.4}^{+15.0}$ & 1.10 \\ 
A2142 offset & 802032010 & 3.46$^\ast$ & 1.41$^\ast$ &  $7.23_{-0.28}^{+0.27}$ &  $0.246_{-0.025}^{+0.042}$ &  $0.607_{-0.19}^{+0.2}$ &  $0.0756_{-0.021}^{+0.026}$ &  $9.79_{-7.1}^{+40.0}$ & 1.22 \\ 
    \end{tabular}
}
     \scalebox{1.0}[1.0]{

    \begin{tabular}{lcccccccc}
      \hline\hline
    \multicolumn{9}{c}{Target (OBSID 805030010) } \\
    \hline 
    \multicolumn{9}{c}{fixed parameters :  $N_\mathrm{H}=3.06$ $\,^\mathrm{a}$, $\Gamma=1.41, S_{\mathrm{CXB}}=6.38 \,^\mathrm{b}, kT_1=0.286$ keV, $kT_2=0.087$ keV  } \\
     Region & $\Omega$ (arcmin$^2$) & $kT_g$ (keV) & $Z_g$ ($Z_\odot$) $\,^\mathrm{d}$ & norm$_g$ $\,^\mathrm{c}$ & norm$_1$ $\,^\mathrm{c}$ & norm$_2$ $\,^\mathrm{c}$ & $\chi^2$/dof & $\sigma_{\mathrm{CXB}}\,^\mathrm{e}$ (\%)  \\
      \hline
$<0.6$ Mpc & 117.4 & $1.64_{-0.11}^{+0.35} \,\pm 0.05 \,^\ddagger$ &  $0.198_{-0.061}^{+0.077} \, \pm 0.01 \,^\ddagger$ &  $6.8_{-1.3}^{+0.8}$ &  $0.62_{-0.14}^{+0.21}$ &  $7.2_{-1.6}^{+1.6}$ & 1.21 & 6 \\ 
$<0.2$ Mpc & 14.1 & $2.43_{-0.36}^{+0.32} \,\pm 0.16 \,^\ddagger$ &  $0.54_{-0.23}^{+0.38}  \,\pm 0.04 \,^\ddagger$ &  $18.2_{-3.0}^{+3.1}$ &  $0.89_{-0.46}^{+0.45}$ &  $7.7_{-5.8}^{+5.8}$ & 1.23 & 18 \\
0.2-0.4 Mpc & 36.1 &  $1.63_{-0.18}^{+0.44} \pm 0.07  \,^\ddagger$ &  $0.22_{-0.10}^{+0.29} \pm 0.02 \,^\ddagger$ &  $7.1_{-1.8}^{+1.4}$ &  $0.49_{-0.24}^{+0.25}$ &  $7.5_{-3.6}^{+3.5}$ & 0.914 & 11 \\ 
0.4-0.6 Mpc & 67.3 &  $1.22_{-0.22}^{+0.30} \pm 0.12 \,^\ddagger$ &  $0.09_{-0.06}^{+0.08} \pm 0.007 \,^\ddagger$ &  $4.0_{-1.1}^{+1.7}$ &  $0.66_{-0.23}^{+0.20}$ &  $6.1_{-2.1}^{+2.1}$ & 1.21 & 8\\ 
hot spot & 14.7 & $3.91_{-0.75}^{+1.06} \pm 0.28 \,^\ddagger$  &  $0.44_{-0.38}^{+0.87} \pm 0.03 \,^\ddagger$ &  $13.9_{-2.9}^{+2.5}$ &  $ <0.62$ &  $ <7.18$ & 1.50 & 17 \\
      \hline
{\hbox to 0pt{\parbox{200mm}{\footnotesize
The units are as follows, $\mathrm{a}: \, 10^{20} \mathrm{cm^{-2}}$, $\mathrm{b}: 10^{-8} \, \mathrm{erg \, cm^{-2} s^{-1} sr^{-1} keV^{-1}}$ for the energy range of 2-10 keV, and  \\ $\mathrm{c}:$ the normalization of the APEC model, norm=$ \int n_e n_\mathrm{H} d V / [4 \pi D_A^2 (1+z)^2]  \times 10^{-20} / \Omega \, \mathrm{cm^{-5} \mathrm{deg}^{-2}} $. \\ 
$\mathrm{d}$ : We use the solar abundance table of \cite{1989GeCoA..53..197A}.
$\mathrm{e}$ : We assume the flux limit of SUZAKU as $10^{-14} \mathrm{ erg/cm^2/s}$. \\ $\ast$ : The parameter is fixed. $\dagger$ : We assume the solar metallically given by \cite{1989GeCoA..53..197A}. \\ $\ddagger$ : The former error indicates the statistical and the latter is the systematics from the temperatures of the galactic components and \\ the CXB.  }\hss}}
    \end{tabular}
}
    \end{center}
\end{table*}

\begin{figure}
\includegraphics[width=\linewidth]{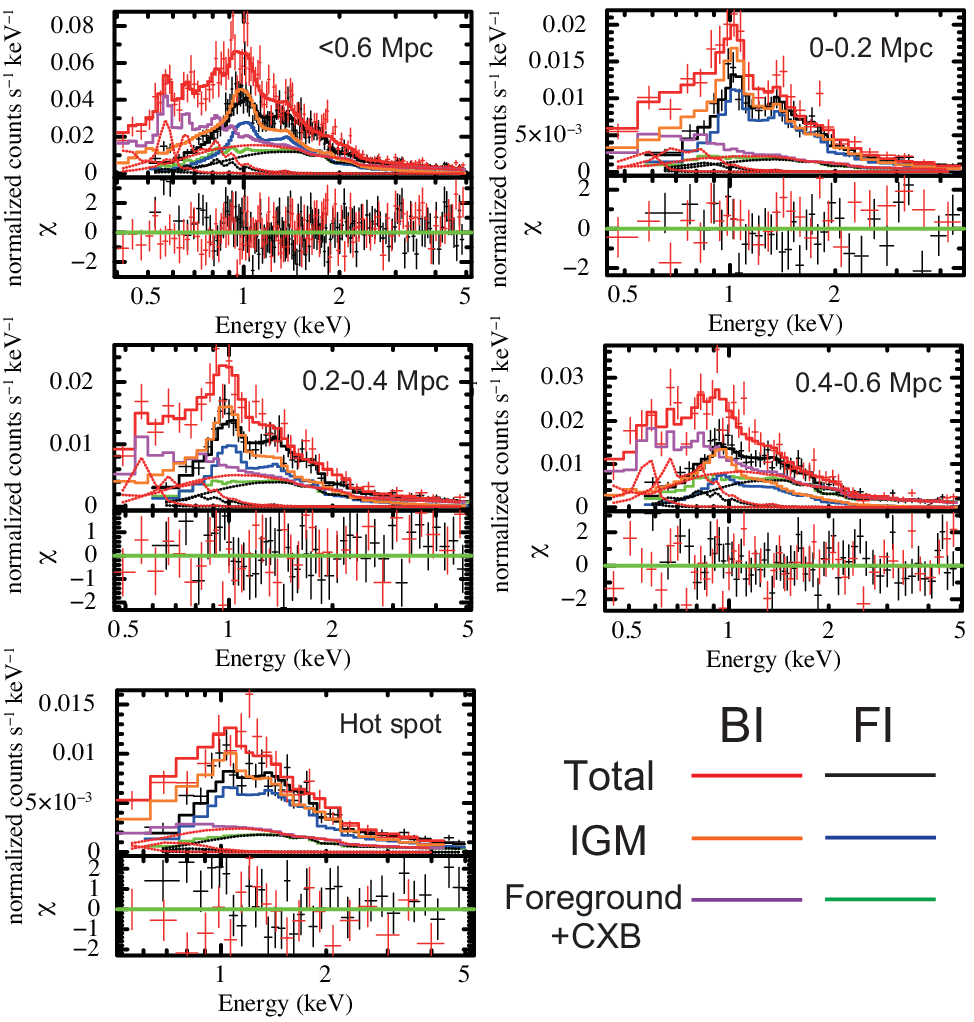}
\caption{Spectra of extracted regions. Each color indicates a different component of spectra: entire spectra (black/red for FI/BI), intragroup gas  (blue/orange for FI/BI), and the foreground and background  (green/magenta) for FI/BI). \label{fig:spec}}
\end{figure}

We extract the spectra of the overall region of the group within the radius of 0.6 Mpc $=6.6^\prime$ around the BCG. The hot spot was excluded with the radius of $0.2 \, \mathrm{Mpc}$ based on the hardness ratio map (Fig [\ref{fig:HR}] c). We fit the spectra by the thermal emission model (${\it apec_g}$) with the two temperature model provided by the offset observation: {\it phabs $\ast$ }({\it pow$+apec_1+apec_g$} ){\it $+apec_2$}. We use the fiducial offset (TCRB) for the temperatures of the galactic components and leave normalizations as free parameters and fix $\Gamma=1.41$ and $S_\mathrm{CXB,asca}$ for the normalization of the CXB. We compute the systematics originated from the uncertainty of the CXB intensity by extrapolating the results of GINGA with the field of views $\Omega$ \citep{2010PASJ...62..371H} and the temperatures of the galactic components assuming that the errors are $\pm 0.05 $ keV and $\pm 0.02$ keV for the TAE and the local component. Although it is difficult to estimate the real fluctuations, the results of the other offset observations are within the assumed errors. The spectra with the fitted lines are shown in Figure \ref{fig:spec} and the best-fit parameters are summarized in Table \ref{tab:spec}.  The spectroscopic temperature is $1.6_{-0.1}^{+0.4} $ keV and the corresponding $R_{500}$ is $0.59 \, \mathrm{Mpc}$ using \citet{2006ApJ...640..691V}, which approximately corresponds to the overall region. The group probably escaped from the detection of the RASS due to its marginal flux ($3.5^{+0.3}_{-0.2} \times 10^{-13} \, \mathrm{erg/s/cm^2}$ for 0.5-2.0 keV) compared with the RASS flux limit $\sim 3 \times 10^{-13}  \, \mathrm{erg/s/cm^2}$. The bolometric luminosity $2.4 \times 10^{43} \, \mathrm{erg/s}$ is consistent with the observed $L_\mathrm{X}$-$T$ relation for $T=1.6$ keV \citep[e.g.][]{2000ARA&A..38..289M}.

As shown in Figure \ref{fig:HR}c, we divide three shells around the BCG center to $R=0.6$ Mpc with the width of $0.2 \mathrm{Mpc} = 2.2^\prime$. The same analysis was done for the three shells and the hot spot. We exclude the hot spot region in the analysis of the three shells. The fitting temperatures and metalicities are provided in Table \ref{tab:spec}. The half decrement of temperature from the center to $R_{500}$ and the abundance profile agree well with a previous systematic study of groups \citep{2007MNRAS.380.1554R,2009ApJ...693.1142S}. The hot spot has significantly higher temperature ($T \sim 4$ keV) than that of surrounding shells, even than the central temperature within $R=0.2$ Mpc and relatively high metalicity which is comparable to that of the center. 

The main systematic source of $\chi^2$ is the slight inconsistency of the BI and FI chips below $E < 0.8$ keV likely due to the Non X-ray Background (NXB) subtraction. However we confirmed this effect does not change the result by neglecting $E < 0.8$ keV. We believe that future update of the NXB calibration will improve this inconsistency.

\section{Discussion} 

The observation of a group with several X-ray peaks is not so common so far. For instance, \citet{2003ApJS..145...39M} compiled 109 nearby groups observed by ROSAT and found only 5 groups have bimodal X-ray peaks and others have rather a single peak.  While \NAME has clear three peaks at least. Moreover, as far as we know, this is the first clear example of the merging group-scale halo ( $T \lesssim 2 $ keV ) with the hot structure associated with the second peak. 

All the second peaks in the bimordal groups of ROSAT have a luminous early-type galaxy as well as the first one. In our target, a relatively faint galaxy is located at the peak B ($z=0.087$; $\times$ in Fig. [\ref{fig:HR}] a) although the physical association is difficult to establish.  Its absolute magnitude in r-band is -21.1, which is not the second brightest but the 9-th brightest one within $R_{500}$, and its color and morphology are red (the rest frame g-r=0.88) and S0 \cite[visually no spiral and the inverse concentration index = 0.37; ][]{2001aj....122.1238S}. 

We confirm that no QSO, group and cluster can be found within $R=10^\prime$ in the SDSS spectroscopic data. We also examined the radio emission in the archival data of NRAO VLA Sky Survey \citep[1.4 GHz][]{1998AJ....115.1693C} and the VLA FIRST survey \citep[21 cm][]{1995ApJ...450..559B} and found two signals in both data within $R=10^\prime$: one is clearly associated with the BCG  and the other is located at the background galaxies ($z \sim 0.14$; indicated by "bg" in Figure \ref{fig:HR} a ). Thus it is unlikely that the peaks B and C are associated with background or foreground objects. 

Though we cannot state any strong conclusions about the detailed merger process due to the limited angular resolution, from the position of the hot spot, the absence of luminous galaxies in the peak B and relative high metalicity in the hot spot, we suggest the following possible scenario: a substructure recently passed close to the BCG with mixing, from north to south, and the central part of the substructure became the brightest peak B by compression, like the bullet cluster \citep{2002ApJ...567L..27M}.

In the above, the presence of the peak C is not necessary. Such a faint peak aligned with two other peaks also has been found in other merging clusters \citep{2002ApJ...565..867S, 2010arXiv1009.1967M}. \cite{2002ApJ...565..867S} discussed the origin of the faint peak in A2256 and suggest an another previous merger origin or the scenario that a fainter peak is originated from the gas of a middle peak, which lagged behind when merger (in this scenario the peak B should move from south). The former scenarios is possible for our target, but the latter do not explain the hot spot. In addition, we propose an another possible scenario : the peaks B and C belonged to one clump before the merging and it fell and split in two. One passed near the center decreasing its velocity by pressure of ambient dense gas (peak B) and the other passed through relatively far from the center or in less dense region (peak C). Thus the peak C went ahead of the peak B. For the last scenario, the merger should be so significant as to separate gas from dark matter subclump. If the galaxy on the peak B is associated with the peak, however, the merger is not so strong. In either case, we suggest that the hotspot is originated from adiabatic compression and heating by the densest clump (peak B).

\begin{figure}[!tbh]
\includegraphics[height=50mm]{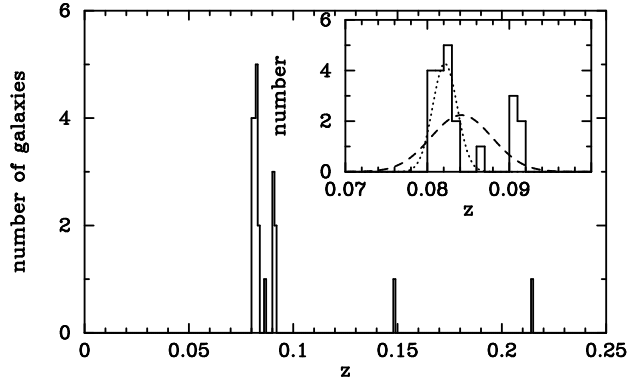}
  \caption{Galaxy distribution of the SDSS spectroscopic data within $R_{500}$. The distribution around the BCG is displayed in the inset. Dashed and dotted curves are fitted Gaussians by the Gapper algorithm using data in  $0.080-0.092$ (both peaks) and $z=0.080-0.088$ (a single peak). The former case fails in the Anderson-Darling test of Gaussianity, while, the latter passes the test. \label{fig:galdi}}
\end{figure}

 Figure \ref{fig:galdi} shows the galaxy distribution of the SDSS spectroscopic data along the line of sight. We found the bimodal distribution around the group. As shown by triangles in Figure \ref{fig:HR} a), the galaxies in the distant peak ($z=0.09-0.092$) are  widely distributed in the group. If we assume both peaks are associated with the group, the velocity dispersion $\sigma_\mathrm{V} = 1197 \pm 191 \, \mathrm{km \, s^{-1}}$ is unusually higher than the expected value from $T$-$\sigma_\mathrm{V}$ and $L_\mathrm{X}$-$\sigma_\mathrm{V}$ relations \citep[e.g.][]{2004MNRAS.350.1511O}. Performing the Anderson-Darling test, we found that the distribution deviates from the Gaussian \citep[$A^{2\ast} = 2.33 $ compared with the critical value of non Gaussian, $A^{2\ast}>$1.93][]{2009ApJ...702.1199H}. If we take the single peak belongs to the target only, $\sigma_V=471 \pm 88 \,\mathrm{km \, s^{-1}}$, which is consistent with the scale relation, and the Gaussian assumption is acceptable ($A^{2\ast} = 0.59 $). If taking the former interpretation, the galaxy distribution supports the merger scenario, while the selection of member galaxies admits of an interpretation. 

The X-ray luminosity is consistent with the L-T relation despite the merging signature. 
It is known that the temperature decrement and density increment in the center of group are shallower than that of massive clusters \citep[e.g.][]{2004MNRAS.350.1511O,2007MNRAS.380.1554R}. This feature might weaken the influence of merger to the luminosity and keep trajectory in the L-T plane to be parallel to the L-T relation during a merger rather than scattering off the relation along either the L or T axis \citep{2001apj...561..621R}. In addition, the large scatter of the L-T relation in the group regime may also help to keep this merging group consistent with the L-T relation.

A connection between the filamentary junction and the merging halo is also an interesting issue although it is difficult to state something from one case. Indeed, most clusters found at the filamentary junction have merging feature, such as multiple peaks, irregular morphology, and radio haloes \citep{2000A&A...355..461A,2004A&A...416..839B,2004A&A...425..429C,2007A&A...470..425B,2008A&A...491..379G}. To explore this connection, we submitted a proposal for further observations of junctions with SUZAKU. 


\acknowledgments 
We are grateful to Noriko Yamasaki, Takaya Ohashi, Yoshitaka Ishisaki, Christophe Pichon, Klaus Dolag, and Eugene Churazov for insightful discussion. We also thank an anonymous referee for a lot of constructive comments. HK is supported by a JSPS Grant-in-Aid for science fellows. This work is supported by Grant-in-Aid for Scientific research from JSPS and from the Japanese Ministry of Education, Culture, Sports, Science and Technology (Nos. 22$\cdot$5467,09J08405, 22-181, and P08324) and by World Premier International Research Center Initiative, MEXT, Japan.

 Funding for the SDSS has been provided by the Alfred P. Sloan Foundation, the Participating Institutions, the National Aeronautics and Space Administration, the National Science Foundation, the U.S. Department of Energy, the Japanese Monbukagakusho, and the Max Planck Society. The SDSS Web site is http://www.sdss.org/.
 
 The SDSS is managed by the Astrophysical Research Consortium for the Participating Institutions. The Participating Institutions are The University of Chicago, Fermilab, the Institute for Advanced Study, the Japan Participation Group, The Johns Hopkins University, Los Alamos National Laboratory, the Max-Planck-Institute for Astronomy, the Max-Planck-Institute for Astrophysics, New Mexico State University, University of Pittsburgh, Princeton University, the United States Naval Observatory, and the University of Washington.

The DSS were produced at the Space Telescope Science Institute under U.S. Government grant NAG W-2166. The images of these surveys are based on photographic data obtained using the Oschin Schmidt Telescope on Palomar Mountain and the UK Schmidt Telescope. The plates were processed into the present compressed digital form with the permission of these institutions.

\end{document}